\documentclass{sigchi}

\toappear{
Permission to make digital or hard copies of part or all of this work for personal or classroom use is granted without fee provided that copies are not made or distributed for profit or commercial advantage and that copies bear this notice and the full citation on the first page. Copyrights for third-party components of this work must be honored. For all other uses, contact the Owner/Author.\\
Copyright is held by the owner/author(s).\\
\emph{CHI 2017}, May 06--11, 2017, Denver, CO, USA\\
ACM 978-1-4503-4655-9/17/05.\\
DOI: http://dx.doi.org/10.1145/3025453.3025576\\
}

\usepackage{graphics}
\usepackage[T1]{fontenc}  
\usepackage{txfonts}
\usepackage{mathptmx}
\usepackage[pdflang={en-US}]{hyperref}
\usepackage{color}
\usepackage{booktabs}
\usepackage{textcomp}
\usepackage{microtype}
\usepackage[all]{hypcap}
\usepackage{ccicons}
\usepackage{todonotes}
\usepackage{verbatim}
\usepackage{balance}

\def\plaintitle{Inferring Cognitive Models from Data \\ using Approximate Bayesian Computation}

\def\plainauthor{Antti Kangasr{\"a}{\"a}si{\"o}, Kumaripaba Athukorala, Andrew Howes, Jukka Corander, Samuel Kaski, Antti Oulasvirta}
\def\plainkeywords{Approximate Bayesian computation;
Cognitive models in HCI; 
Computational rationality;
Inverse modeling}

\makeatletter
\def\url@leostyle{%
  \@ifundefined{selectfont}{
    \def\UrlFont{\sf}
  }{
    \def\UrlFont{\small\bf\ttfamily}
  }}
\makeatother
\def\pprw{8.5in}
\def\pprh{11in}

\definecolor{linkColor}{RGB}{6,125,233}
\hypersetup{%
  pdftitle={\plaintitle},
  pdfauthor={\plainauthor},
  pdfkeywords={\plainkeywords},
  pdfdisplaydoctitle=true, 
  bookmarksnumbered,
  pdfstartview={FitH},
  colorlinks,
  citecolor=black,
  filecolor=black,
  linkcolor=black,
  urlcolor=linkColor,
  breaklinks=true,
  hypertexnames=false
}

\newcommand\tabhead[1]{\small\textbf{#1}}

\begin{document}

\title{\plaintitle}

\numberofauthors{6}
\author{%
Antti Kangasr{\"a}{\"a}si{\"o}$^{1}$, Kumaripaba Athukorala$^{1}$, Andrew Howes$^{2}$,\\
Jukka Corander$^{3}$, Samuel Kaski$^{1}$, Antti Oulasvirta$^{4}$\\
\affaddr{$^1$Helsinki Institute for Information Technology HIIT,} \\
\affaddr{Department of Computer Science, Aalto University, Finland} \\
\affaddr{$^2$School of Computer Science, University of Birmingham, UK} \\ 
\affaddr{$^3$Department of Biostatistics, University of Oslo, Norway} \\
\affaddr{$^4$Helsinki Institute for Information Technology HIIT,} \\
\affaddr{Department of Communications and Networking, Aalto University, Finland}
}

\maketitle

\begin{abstract}
An important problem for HCI researchers is to estimate the parameter values of a cognitive model from behavioral data.
This is a difficult problem, because of the substantial complexity and variety in human behavioral strategies. 
We report an investigation into a new approach using approximate Bayesian computation (ABC) to condition model parameters to data and prior knowledge.
As the case study we examine menu interaction, 
where we have click time data only to infer a cognitive model that implements a search behaviour with parameters such as fixation duration and recall probability.
Our results demonstrate that ABC
(i) improves estimates of model parameter values,
(ii) enables meaningful comparisons between model variants, and
(iii) supports fitting models to individual users.
ABC provides ample opportunities for theoretical HCI research by allowing principled inference of model parameter values and their uncertainty.
\end{abstract}



\section{Introduction}

It has become relatively easy to collect large amounts of data about complex user behaviour.
This provides an exciting opportunity as the data has the potential to help HCI researchers understand and possibly predict such user behavior.
Yet, unfortunately it has remained difficult to explain what users are doing and why in a given data set.

\begin{figure}[!t]
\centering
\includegraphics[width=0.85 \columnwidth,natwidth=2000,natheight=1376]{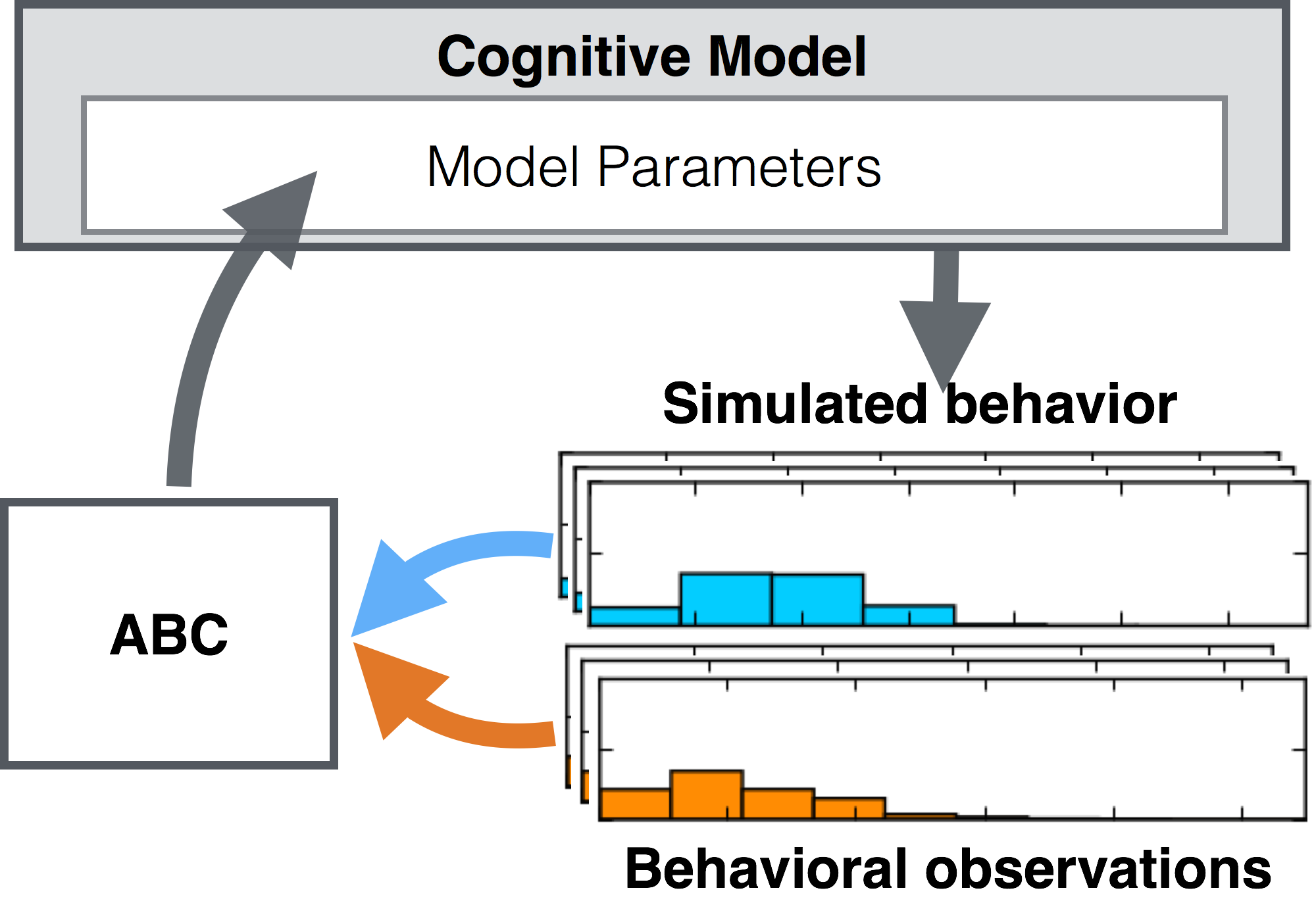}
\caption{
This paper studies methodology for inference of parameter values of cognitive models from observational data in HCI. 
At the bottom of the figure, we have behavioral data (orange histograms), such as times and targets of menu selections.
At the top of the figure, a cognitive model generates simulated interaction data (blue histograms). 
In this paper, approximate Bayesian computation (ABC) is investigated to identify the model parameter values that yield the best fit between the real data and simulator-generated data, while keeping the parameter values reasonable given prior knowledge.
}
\label{fig:overview1}
\end{figure}

The difficulty lies in two problems: modeling and inference.
The \emph{modeling problem} consists of building models that are sufficiently general to capture a broad range of behaviors.
Any model attempting to explain real-world observations must cover a complex interplay of factors, including what users are interested in, their individual capacities, and how they choose to process information (strategies). 
Recent research has shown progress in the direction of creating models for complex behavior 
\cite{bailly2014model,
chen2015emergence,
Cockburn2007,
fu2007snif,
Halverson2011,
Hornof2004,
Kieras2014,
kieras2000modern,
miller2004,
payne2013adaptive}.
After constructing the model, we are then faced with the \emph{inference problem}: how to set the parameter values of the model, such that the values agree with literature and prior knowledge, and that the resulting predictions match with the observations we have  (Figure~\ref{fig:overview1}).
Unfortunately, this problem has been less systematically studied in HCI.
To this end, the goal of this paper is to report an investigation into a flexible and powerful method for inferring model parameter values, called \emph{approximate Bayesian computation} (ABC) \cite{sunnaaker2013approximate}.

ABC has been applied to many scientific problems \cite{beck2004model,csillery2010approximate,sunnaaker2013approximate}.
For example, in climatology the goal is to infer a model of climate from sensor readings, and in infectious disease epidemiology an epidemic model from reports of an infection spread.
Inference is of great use both in applications and in theory-formation, in particular when testing models, identifying anomalies, and finding explanations to observations.
However ABC, nor any other principled inference method, have, to our knowledge, been applied to complex cognitive models in HCI\footnote{For simpler models, such as regression models (e.g., Fitts' law), there exist well-known methods for finding parameter values, such as ordinary least squares.}.

We are interested in principled methods for inferring parameter values, because they would be especially useful for process models of behaviour.
This is because the models are usually defined as  simulators, and thus the inference is very difficult to perform using direct analytical means\footnote{In technical terms, such models generally do not have a \emph{likelihood function}---defining the likelihood of parameter values given the observations---that could be written in closed form.}.
Such process models in HCI have been created, for example, based on cognitive science \cite{anderson1997act,byrne2001act,card1983psychology,fu2007snif,kieras1997overview,rumelhart1982simulating}, control theory \cite{jagacinski2003control}, biomechanics \cite{bachynskyi2015informing}, game theory \cite{camerer2003behavioral}, foraging  \cite{pirolli1999information,pirolli2005rational}, economic choice \cite{azzopardi2014modelling}, and computational rationality \cite{chen2015emergence}.
In the absence of principled inference methods for such models,
some approaches have included:
(1) simplifying models until traditional inference methods are possible;
(2) using values adopted from the literature or adjusting them without studying their effect on behavior;
or (3) manually iterating to find values that lead to acceptable performance.
Compared to this, principled inference methods might reduce the potential for ambiguity, miscalculation, and bias,
because model parameter values could be properly conditioned on both literature and prior knowledge, as well as the observation data.

ABC is particularly promising for inferring the values of process model parameters from naturalistic data---a problem that is known to be difficult in cognitive science \cite{myung2016model}.
The reason is that ABC does not make any further assumptions of the model, apart from the researcher being able to repeatedly simulate data from it using different parameter values.
ABC performs inference by systematically simulating user behavior with different parameter configurations.
Based on the simulations, ABC estimates which parameter values lead to behavior that is similar to observations, while also being reasonable considering our prior knowledge of plausible parameter values.

As a challenging and representative example, this paper looks at a recent HCI process model class in which behavioral strategies are learned using reinforcement learning 
\cite{chen2015emergence,fu2007snif,gershman2015computational,payne2013adaptive}.
These models assume that users behave (approximately) to maximize utility given limits on their own capacity.
The models predict how a user will behave in situations constrained by
(1) the environment, such as the physical structure of a user interface (UI);
(2) goals, such as the trade-off between time and effort;
and (3) the user's cognitive and perceptual capabilities, such as memory capacity or fixation duration.
This class of models, called \emph{computational rationality} (CR) models, has been explored previously in HCI, for example in SNIF-ACT \cite{fu2007snif}, economic models of search \cite{azzopardi2014modelling}, foraging theory \cite{pirolli1999information}, and adaptive interaction \cite{payne2013adaptive}.
The recent interest in this class is due to the benefit that, when compared with classic cognitive models, it requires no predefined specification of the user's task solution, only the objectives.
Given those, and the constraints of the situation, we can use machine learning to infer the optimal behavior policy. 
However, achieving the inverse, that is inferring the constraints assuming that the behavior is optimal, is exceedingly difficult.
The assumptions about data quality and granularity of previously explored methods for this inverse reinforcement learning problem \cite{ng2000algorithms,ramachandran2007bayesian,ziebart2008maximum} tend to be unreasonable when often only noisy or aggregate-level data exists, such as is often the case in HCI studies.

Our application case is a recent model of \emph{menu interaction} \cite{chen2015emergence}.
The model studied here has previously captured adaptation of search behavior, and consequently changes to task completion times, in various situations \cite{chen2015emergence}.
The model makes parametric assumptions about the user, for example about the visual system (e.g., fixation durations), and uses reinforcement learning to obtain a behavioral strategy suitable for a particular menu. 
The inverse problem we study is how to obtain estimates of the properties of the user's visual system from selection time data only (click times of menu items).
However, due to the complexity of the model, its parameter values were originally tuned based on literature.
Later in Study 1, we demonstrate that we are able to infer the parameter values of this model based on observation data, such that the predictions improve over the baseline, while the parameter values still agree with the literature.
To the best of our knowledge, this is also the first time such inverse reinforcement learning problem has been solved based on aggregate-level data.

We also aim to demonstrate the applicability of ABC, and inference in general, in two situations: model development and modeling of individuals.
In Study 2, we demonstrate how ABC allows us to make meaningful comparisons between multiple model variants, and their comparable parameters, after they all have been fit to the same dataset.
This presents a method for speeding up the development of these kind of complex models, though automatic inference of model parameter values.
In Study 3, we demonstrate how ABC allows us to infer model parameter values for individual users.
We discover that overall these individual models outperform a population-level model fit to a larger set of data, thus demonstrating the benefit of individual models.
As a comparison, it would not be possible to fit individual models based on literature alone, as the information generally only applies on population level.

\section{Overview of Approach}

This paper is concerned with inference of model parameter values from data, which is also called \emph{inverse modeling}.
Inverse modeling answers the question: ``what were the parameter values of the model, assuming the observed data was generated from the model?''
Our goal is to assess the usefulness of approximate Bayesian computation (ABC) \cite{sunnaaker2013approximate} to this end.

We now give a short overview of inverse modeling in HCI, 
after which we review ABC and explain its applicability. 
We finally provide a short overview of the particular ABC algorithm, BOLFI \cite{gutmann2016bayesian}, we use in this study.

\subsection{Inverse Modeling Approaches for Cognitive Models}

For models that have simple algebraic forms, such as linear regression, inverse modeling is simple, as we can explicitly write down the formula for the most likely parameter values given data.
For complex models, such formula might not exist, but it is often possible to write down an explicit \emph{likelihood function}, $L(\theta|Y_{obs})$, which evaluates the likelihood of the parameters $\theta$ given the observed data $Y_{obs}$.
When this likelihood function can be evaluated efficiently, inverse modeling can be done, even for reinforcement learning (RL) models \cite{ng2000algorithms,ramachandran2007bayesian,ziebart2008maximum}.
However, this inverse reinforcement learning has been only possible when precise observations are available of the environment states and of the actions the agent took, which in HCI applications is rarely the case.

When the likelihood function of the model can not be evaluated efficiently, there are generally two options left.
The traditional way in HCI has been to set the model parameters based on past models and existing literature.
If this has not led to acceptable predicted behavior, the researcher might have further tuned the parameters by hand until the predictions were satisfactory.
However, this process generally has no guarantees that the final parameters will be close to the most likely values.
An alternative solution, which we have not seen used in HCI context before, would be to use likelihood-free inference methods, that allow the model parameters to be estimated without requiring the likelihood function to be evaluated directly.
These methods are derived based on mathematical principles, and thus offer performance guarantees, at least in the asymptotic case.
ABC is one such method \cite{sunnaaker2013approximate}, and we will explain it next in more detail.

\begin{figure}[!t]
\centering
\includegraphics[width=0.85 \columnwidth,natwidth=899,natheight=301]{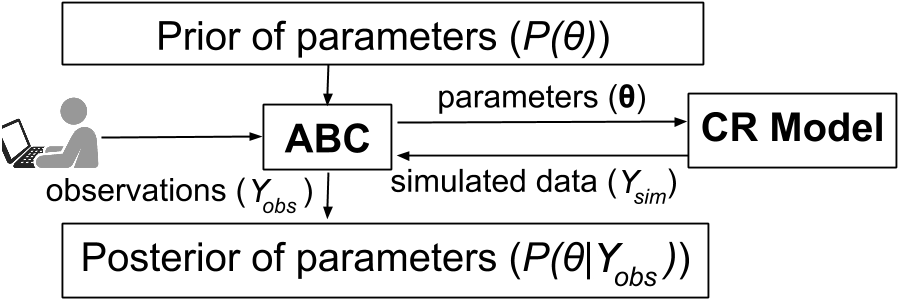}
\caption{
Overview of the ABC inference process for HCI models: Observed user data and priors of the parameters are fed into the ABC algorithm, which then approximates the posterior distribution of the parameter values.
The algorithm iterates by choosing values for the parameters of the model (here a CR model) and generating simulated user data. For CR models, generating simulated data requires first training a reinforcement learning agent using the given parameter values.
}
\label{fig:improcess}
\end{figure}

\subsection{Approximate Bayesian Computation (ABC)}

ABC is a principled method for finding parameter values for complex HCI models, including simulators, based on observed data and prior knowledge.
It repeatedly simulates data using different parameter values, in order to find regions of the parameter space that lead to simulated data that is similar to the observed data.
Different ABC algorithms differ, for example, in the way in which they choose the parameter values.

The main benefit of ABC for HCI is its generality: the only assumption needed is that the researcher is able to repeatedly simulate observations with different parameter values.
Therefore, while in this paper we examine only a particular simulator, the approach is of more general value.
To be precise, ABC can be used in the following recurring scenario in HCI:
\begin{itemize}
\item \textbf{Inputs:} A model $M$ with unknown parameters $\theta$; prior knowledge of reasonable values for $\theta$ (for example from literature); observations $Y_{obs}$ of interactive behavior (for example from user study logs)
\item \textbf{Outputs:} Estimates of likely values for parameters $\theta$ and their uncertainty. Likely values of $\theta$ should produce a close simulated replication of observed data: $M(\theta) \approx Y_{obs}$, while still being plausible given prior knowledge.
\end{itemize}

The process of using ABC is depicted in Figure \ref{fig:improcess}.
First the researcher implements her model as an executable simulator.
Values for well-known parameters of the model are set by hand.
For inferred parameters $\theta$ a \emph{prior probability distribution} $P(\theta)$ is defined by the researcher based on her prior knowledge of plausible values.
The researcher then defines the set of observations $Y_{obs}$ that $\theta$ will be conditioned on.
Next, the researcher defines a \emph{discrepancy function} $d(Y_{obs}, Y_{sim}) \to [0,\infty)$, that quantifies the similarity of the observed and simulated data in a way meaningful for the researcher.
Finally, an ABC algorithm is run; it selects at which parameter values $\{\theta_i\}$ the simulator will be run, and how the conditional distribution of the parameter values, also known as the \emph{posterior} $P(\theta|Y_{obs})$, is constructed based on the simulations.

\subsection{BOLFI: An ABC Variant Used in This Paper}

\begin{figure}[!b]
\centering
\includegraphics[width=1.0 \columnwidth,natwidth=648,natheight=218]{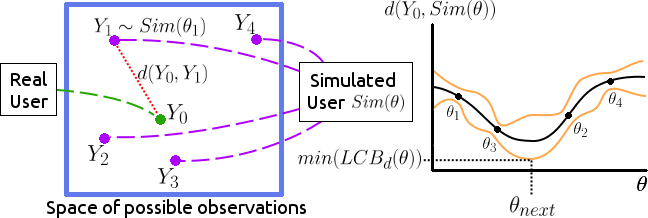}
\caption{
Left: BOLFI finds parameter values that are best able to reproduce empirical observations $Y_0$ (here the best sample is $Y_3$, produced by simulating with parameter values $\theta_3$).
Right: BOLFI first constructs a statistical regression model for predicting the discrepancy values $d$ associated with different parameter values $\theta$, and then uses Lower Confidence Bound (LCB) values for choosing the next sample location $\theta_{next}$.
}
\label{fig:overview2}
\end{figure}

This paper employs a recent variant of ABC called BOLFI \cite{gutmann2016bayesian}, which reduces the number of simulations\footnote{The naive way to use ABC would be to simulate a large amount of samples densely covering the parameter space and keep those that have the lowest discrepancy values.
This method is also known as Rejection ABC.
However, as in our case the simulations take multiple hours each, this approach has infeasible total computation time.} while still being able to get adequate estimates for $\theta$. An overview of the method is shown in Figure~\ref{fig:overview2}.

The main idea of BOLFI is to learn a statistical regression model---called a Gaussian process---for estimating the discrepancy values over the feasible domain of $\theta$ from a smaller number of samples that do not densely cover the whole parameter space.
This is justified when the situation is such that small changes in $\theta$ do not yield large changes in the discrepancy.
Additionally, as we are most interested in finding regions where the discrepancy is small, BOLFI uses a modern optimization method called \emph{Bayesian optimization} for selecting the locations where to simulate.
This way we can concentrate the samples to parameter regions that are more likely to lead to low discrepancy simulated data.
This approach has resulted in 3--4 orders of faster inference compared with the state-of-the-art ABC algorithms.
Details of the method are given in the paper by Gutmann and Corander \cite{gutmann2016bayesian}.

\section{Case: Model of Menu Selection}

\begin{figure}[!b]
\centering
\includegraphics[width=0.9\columnwidth,natwidth=1108,natheight=801]{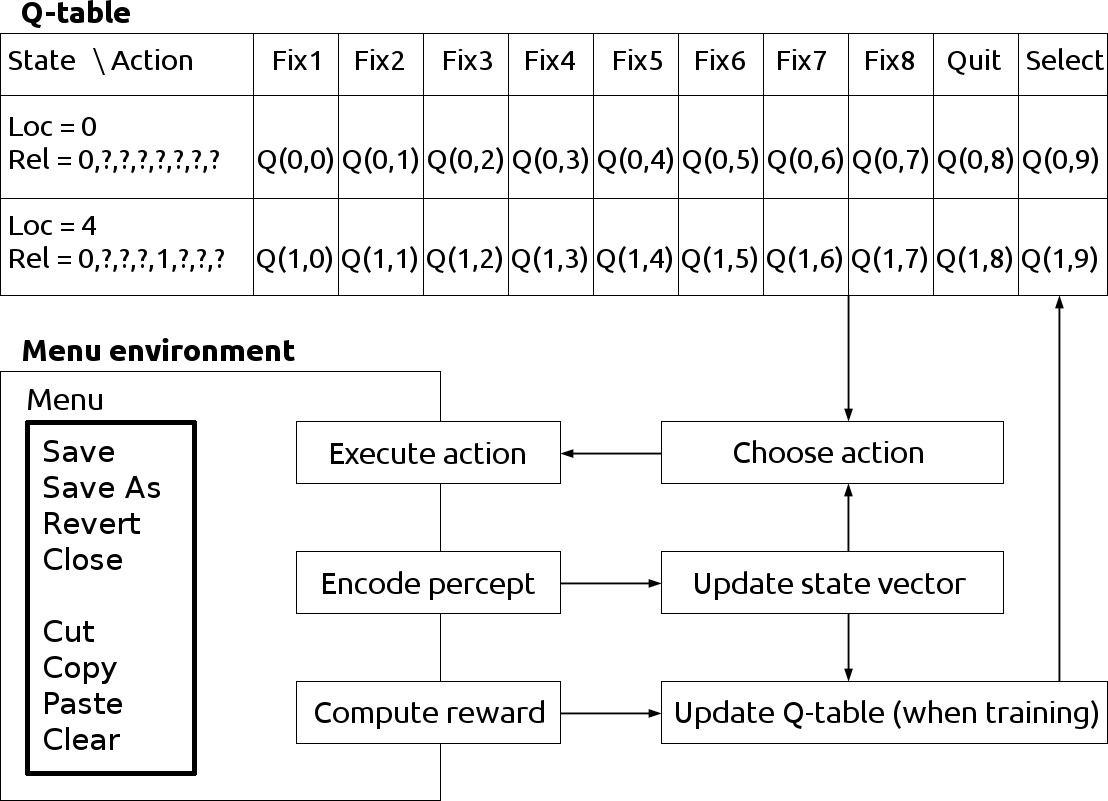}
\caption{
Case model: A simulation process figure of the computational rationality model of user interaction with drop-down menus, adapted from Chen et al. \cite{chen2015emergence}.
The Q-table is constructed in the training phase; in the simulation phase it is kept fixed.
}
\label{fig:CR-menu-model}
\end{figure}

Our case looks at a recent model for visual search of menus, introduced by Chen et al. \cite{chen2015emergence}.
The purpose of this model is to predict the visual search behavior (how eyes fixate and move) and task completion times of a person searching for an item from a vertical menu.

This model presents a particularly challenging problem for inference of parameter values because substantial computation is required by the reinforcement learning algorithm to calculate the search behavior policy, given a particular parameter set.
The parameters of Chen et al.'s model \cite{chen2015emergence} describe cognitive characteristics of a user, such as the duration of a saccade when searching through the menu.
In contrast to Chen et al. \cite{chen2015emergence}, where parameters were largely set to values in the literature\footnote{
 For example, the saccade duration parameters were set based on a study by Baloh et al. \cite{baloh1975quantitative} and the fixation duration parameters based on a study by Brumby et al. \cite{Brumby:2014}.
}, the inference problem that we study here is to estimate parameter values based on limited behavioral data: click times for menu items.
Across the studies, we condition the parameter values of this model, and it's variants, to this type of data in different settings.

\subsection{Introduction to Computational Rationality}

An important property of the model we examine \cite{chen2015emergence} is that it computes computationally rational policies---behavior patterns optimized to maximize the utility of an agent given its bounds and goals \cite{lewis2014computational}.
The bounds include limitations on the observation functions and on the actions that the agent can perform.
These bounds define a space of possible policies.
The use of computationally rational agents to model cognition has been heavily influenced by \emph{rational analysis}, a method for explaining behavior in terms of utility \cite{anderson1991human,chater1999ten,oaksford1994rational},
an idea used for example in information foraging theory and economic models of search \cite{azzopardi2014modelling,pirolli1999information}.
Computational rational agents have been used to model a number of phenomena in HCI \cite{payne2013adaptive}.
Applications relevant to this paper include menu interaction \cite{chen2015emergence} and visual search \cite{hayhoe2014modeling,Myers2013,nunez2013models,Tseng2015}.

CR models use \emph{reinforcement learning} (RL) methods to compute the optimal policies \cite{sutton1998reinforcement}. 
Applying RL has two prerequisites.
First, an environment is needed, which has a state that the RL agent can observe, and actions that the agent can perform to change the state.
The environment is commonly a Markov decision process (MDP), and designed to approximate the real-world situation the real user faces.
Second, a reward function is required---a mapping from the states of the environment to real numbers---which defines what kind of states are valuable for the RL agent (higher rewards being favorable).
The RL algorithm finds the (approximately) optimal policy by experimenting in the environment and updating the policy until (approximate) convergence.
The resulting policy---and thus the predicted behavior---naturally depends on the parameters of the environment and of the reward function, which have been set by the researcher.

\subsection{Overview of Menu Selection Model}

We summarize here all key details of the original model of Chen et al. \cite{chen2015emergence} (Fig \ref{fig:CR-menu-model}).

The environment is a menu composed of eight items, arranged into two semantic groups of four items each, where the items in each group share some semantic similarity.
There are two conditions for the menu: either the item is present in the menu, or absent.
At the beginning of an episode the agent is shown a target item.
The task of the agent is to select the target item in the menu if it is present, or otherwise to declare that the menu does not contain the target item.

The agent has ten possible actions: fixate on any of the 8 items, select the fixated item or declare that the item is not present in the menu (quit).
Fixating on an item reveals its semantic relevance to the agent, whereas selecting an item or quitting ends the episode.
After each action, the agent observes the state of the environment, represented with two variables: semantic relevances of the observed menu items
and the current fixation location.
The agent receives a reward after each action.
After a fixation action, the agent gets a penalty that corresponds to the time spent for performing the saccade from the previous location and the fixation to the new item.
If the agent performs the correct end action, a large reward is given---otherwise an end action results in a large penalty.

The RL agent selects the actions based on the expected cumulative rewards the action allows the agent to receive starting from the current state---also known as Q-value of the state-action pair in RL terminology.
These Q-values are learned in the training phase, over 20 million training episodes, using the Q-learning algorithm.
To select an action, the agent compares the Q-values of each action in that state (see the Q-table in Fig \ref{fig:CR-menu-model}) and chooses the action with the highest value.

\begin{table}[!t]
\centering
\begin{tabular}{p{1.5cm}p{5cm}}
\hline
\tabhead{Parameter} & \tabhead{Description} \\ \hline
        $f_{dur}$ & Fixation duration       \\\hline
        $d_{sel}$ & Time cost for selecting an item (added to the duration of the last fixation of the episode if the user made a selection) \\\hline
        $p_{rec}$ &  Probability of recalling the semantic relevances of all of the menu items during the first fixation of the episode           \\\hline
        $p_{sem}$ &     Probability of perceiving the semantic relevance of menu items above and below of the fixated item          \\\hline
\end{tabular}
\caption{Parameters inferred with ABC in Studies 1-3.}
\label{tab:parameters}
\end{table}

\subsection{Variants}

Above we described one model variant reported in Chen et al. \cite{chen2015emergence}.
According to the description of the observation data, no items in the menus had more than 3 letters difference in length \cite{bailly2014model}. 
To comply with this and to reduce the complexity of the state space, we assumed that there is no detectable difference in the length of the items. 
Thus we used the model variant from Chen et al. \cite{chen2015emergence} where the only detectable feature is the semantic similarity to the target item.
In Study 2 reported below, we will explore three additions to the model and their effect on the predictions.
All model parameters inferred with ABC, across the studies, are listed in Table \ref{tab:parameters}. 

\section{Experiments and Results}

\begin{figure*}[!ht]
\centering
\includegraphics[width=0.85 \textwidth,natwidth=1125,natheight=669]{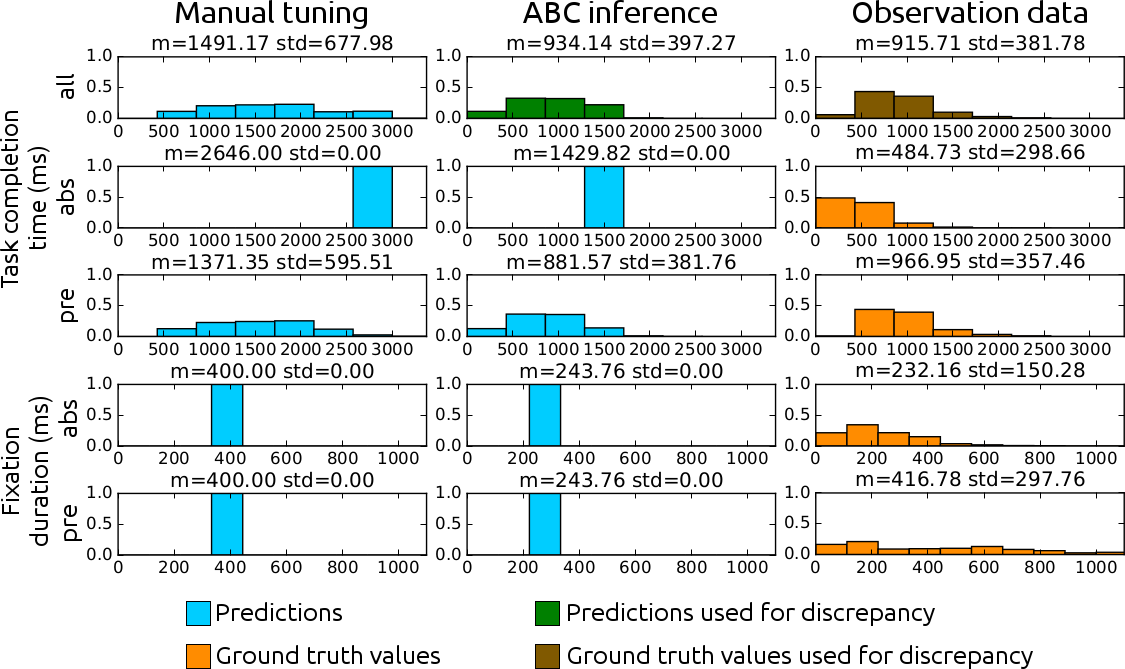}
\caption{
Comparison of manual tuning to ABC inference (Study 1).
Predictions made by conditioning parameter values on both aggregate-level observed data and prior knowledge (ABC inference; blue bars in the middle column) agree better with observation data not seen by the model (orange bars on the right column) than predictions made by setting parameter values manually based on literature (Manual tuning; blue bars in the left column).
ABC searched for parameter values that resulted in small discrepancy between the model predictions (green) and observed aggregate-level data (brown).
\emph{Left column}: Manual tuning: all parameter values were set based on literature and manual tuning.
\emph{Center column}: ABC inference: the value of $f_{dur}$ has been conditioned on observation data using ABC.
\emph{Right column}: Observation data (ground truth) from Bailly et al. \cite{bailly2014model}.
\emph{All}: Aggregated data from both conditions.
\emph{Abs}: Data from when target was absent from the menu.
\emph{Pre}: Data from when target was present in the menu.
}
\label{fig:exp1-results-1}
\end{figure*}

In the rest of the paper, we show with three case studies how ABC can be used to improve the current modeling practices. All studies use the Chen et al. model \cite{chen2015emergence}, and the core problem in all is \emph{inverse modeling}:
Given aggregate observation data (task completion times), find the most likely parameter values $\theta$ and their distribution, such that the predictions made by the model agree with the observations.

\begin{enumerate}
\item \textbf{Study 1. ABC compared to manual tuning}: We demonstrate that ABC can improve model fit by inferring parameter values from data, compared to the common practice of setting them manually based on the literature.
\item \textbf{Study 2. ABC in model development}: We demonstrate how ABC helps in improving models, by fitting multiple models to same data, exposing differences and anomalies.
\item \textbf{Study 3. ABC in modeling individual differences}: We demonstrate how individual models can be fit with ABC, by conditioning the model to individual data.
\end{enumerate}

We use the same dataset as Chen et al. \cite{chen2015emergence}, 
which is a subset of a study reported by Bailly et al. \cite{bailly2014model} and based on the study design of Nilsen \cite{nilsen1999exploring}.
In the study, a label is shown and the user must click the correct item in a menu with 8 elements as quickly as possible.
Items were repeated multiple times to understand practice effects.
Multiple menus were used, and target position and absence/presence of target systematically varied.
Eye movement data were collected and processed for fixation and saccade durations.
Twenty-one paid participants took part in the study.
Further details of the study that produced the data are reported in \cite{bailly2014model}.

We implemented the BOLFI algorithm in Python.
Parts of the source code were later published within an open-source library for likelihood-free inference \cite{kangasraasio2016engine}.
Running the experiments took around one day each on a cluster computer.
Further technical details of the experiments and implementation are described in the Appendix.

\subsection{Study 1. ABC Compared to Manual Tuning}

Our aim in the first study was to analyze how much we can improve the predictions made by the model by conditioning values of key parameters on observation data instead of the standard practice of choosing all of the parameter values manually.
The case study was chosen to represent the common setting in HCI research where only aggregate data may be available.

We used the model of Chen et al. \cite{chen2015emergence}, and compared the parameter values inferred by ABC to those set based on literature in the original paper \cite{chen2015emergence}.
We predicted task completion times (TCT) and fixation durations with both models, and compared them with observation data from \cite{bailly2014model}.
For simplicity, we inferred the value of only one parameter $\theta$ with ABC, the \emph{fixation duration} $f_{dur}$.
The rest of the model parameter values were set to be identical with the baseline model.
The value of this parameter was conditioned on the observed aggregate task completion times (TCT; combined observations from both menu conditions: target absent---referred to as \emph{abs}, target present---referred to as \emph{pre}).
Chen et al. \cite{chen2015emergence} set the value of this parameter to 400~ms based on a study by Brumby et al. \cite{Brumby:2014}.

\subsubsection{Results}

As shown in Figure~\ref{fig:exp1-results-1}, the parameter value inferred with ABC lead to the model predictions matching better to observation data not used for the modelling. 
This holds both for TCT and fixation duration.
In detail, the ground truth aggregated TCT was 0.92~s~(std~0.38~s).
The manually fit model predicted 1.49~s~(std~0.68~s), whereas the ABC fit model predicted 0.93~s~(std~0.40~s).
For predictions, we used the maximum a posteriori (MAP) value predicted by ABC, which was 244~ms for fixation duration (detail not shown). 
This corresponds to values often encountered in e.g. reading tasks \cite{rayner1998eye}.

In summary, inferring the fixation duration parameter value using ABC lead to improved predictions, compared to setting the parameter value manually based on literature.
The inferred parameter value was also reasonable based on literature.

\subsubsection{Observations on the Resulting Models}

A closer inspection of predictions made by the models exposed two problematic issues which led to improvements in Study 2.
The first issue is that while the aggregate TCT predictions were accurate, and all predictions with ABC were better compared to manual tuning, even ABC-fitted predictions were not reasonable when split to sub-cases according to whether the target was present in the menu or not.
This is clearly visible in Figure~\ref{fig:exp1-results-1} (rows two and three), where we notice that the predicted TCT when target is absent is actually around four to six times as long as the actual user behavior. 

\begin{figure}[!tb]
\centering
\includegraphics[width=0.45 \textwidth,natwidth=690,natheight=246]{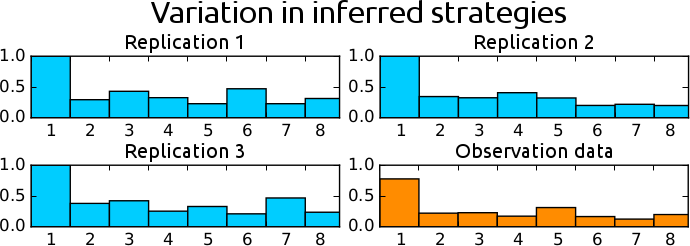}
\caption{
Study 1: Repeated execution of the Chen et al. \cite{chen2015emergence} model yields variations in search patterns.
None of the three independent realizations (Replication 1-3) was able to reproduce the noticeable spike on the 5th item (i.e., the first item of the second semantic group) in the Observation data.
The bar charts illustrate average proportions of \emph{gazes to target} in search episodes, as a function of the target location.
Larger proportions indicate that targets at that location are on average found earlier, as fewer gazes to non-target items are required.
}
\label{fig:exp1-results-3}
\end{figure}

The second issue concerns the search strategies predicted by the model.
Chen et al. \cite{chen2015emergence} showed that their model was able to learn a behavior strategy, where the agent would look first at the topmost item, and second at the 5th item, which was the first item of the second semantic group.
This was seen as a clear spike on the fifth item in the ``proportion of gazes to target'' feature.
However, not every attempt to replicate this result succeeded
(Fig.~\ref{fig:exp1-results-3}), and similar variation in predicted strategies was observed with the ABC-fitted model as well\footnote{
The only technical difference between the original vs. our
implementation was that in the original \cite{chen2015emergence} Q-learning was performed on a predetermined set of 10,000 menu realizations, whereas we generated a new menu for every training session.
The original implementation thus converged slightly faster, as it explored a smaller part of the state space.
}.

Our conclusion is that there likely exist multiple behavior strategies that are almost equally optimal, and the RL algorithm may then find different local optima in different realizations.
This is possible, as Q-learning is guaranteed to find the globally optimal strategy only given an infinite amount of learning samples; with only finite samples, this is not guaranteed.
Because of this issue with the inference of the behavioral strategies, we do not discuss in detail the inferred strategies, but only report results that we were able to repeat reliably.

\subsection{Study 2: ABC in Model Development}

We next demonstrate how ABC can be used in the model improvement cycle, where new models are proposed and compared.
As a baseline, we start with the model introduced in Study 1, to which we add features to fix the issues we observed in Study 1.
We show that with ABC multiple different models can be conditioned to the same observation data, in order to compare their predictions and (compatible) parameter estimates.
Doing the same manually would be very laborious.

The model variants we propose are as follows:
\begin{itemize}
\item \textbf{Variant 1:
Chen et al. \cite{chen2015emergence} model $+$ selection latency}:
The agent incurs a delay $d_{sel}$ when selecting an item.
\item \textbf{Variant 2: Variant 1 $+$ immediate menu recognition}:
The agent is able to recognize the menu based on the first item with probability $p_{recall}$.
\item \textbf{Variant 3: Variant 2 $+$ larger foveated area}:
The agent can perceive the semantic relevance of the neighboring items (above and below the fixated item) through peripheral vision with probability $p_{sem}$.
\end{itemize}

\textbf{Variant 1:}
We first observed that both the TCT and recorded fixation duration are longer when the target item is present.
We hypothesized that the user might have had to spend some time confirming her judgment of the target item and physically making the selection using the pointer.
To allow the model to capture this behavior, we added an additional delay, $d_{sel}$, for the selection action.
For example, the mathematical model of Bailly et al. \cite{bailly2014model} implements a similar selection latency.

\textbf{Variant 2:}
We observed that some of the users were able to decide that the target item was not present in the menu just using one fixation on the menu.
Our hypothesis was that the users were able to memorize some of the menus, allowing them to naturally finish the task much faster when they recalled the menu layout.
To capture this behavior, we allowed the agent to instantly observe the full menu during the first fixation, with probability $p_{rec}$.

\textbf{Variant 3:}
We also observed in Study 1 that the inferred number of fixations was in both cases larger than in the observation data.
The models predicted on average 6.0 fixations when the target was absent (ground truth was 1.9) and 3.1 when target was present (ground truth was 2.2).
Our hypothesis was that the user might have observed the semantic relevance of neighboring items using peripheral vision, allowing her to finish the task with a smaller number of fixations.
The model of Chen et al. \cite{chen2015emergence} had a peripheral vision component but it only applied to size-related information (shape relevance).
Our hypothesis is also justified by the experiment setup of Bailly et al. \cite{bailly2014model}, where the neighboring items do fall within the fovea (2 degrees), thus making it physiologically possible for the user to observe the semantic relevance of the neighboring items.
To capture this behavior, we allowed the agent to observe the semantic relevance of the items above and below the fixated item in the menu with probability $p_{sem}$ (independently for the item above and below).

\emph{Implementation:}
In order to be able to do inference on these new parameters, we only needed to make small additions to the simulator code: add an interface for setting the values of these new parameters and implement the described changes in the model.
On the ABC side, we only described the names and priors of the new parameters, and increased the amount of locations where to simulate.
More locations are justified as each new parameter increases the size of the parameter space that needs to be searched.
We also noticed in Study 1 that the models were not able to replicate the behavior well in both menu conditions (target present, absent) at the same time.
For this reason, we make a small adjustment to the discrepancy function, so that the TCT is compared in both menu conditions separately.
This should allow the models to better replicate the full observed behavior.
Further details are provided in the Appendix.

\subsubsection{Results}

\begin{figure*}[!ht]
\centering
\includegraphics[width=0.98 \textwidth,natwidth=1824,natheight=683]{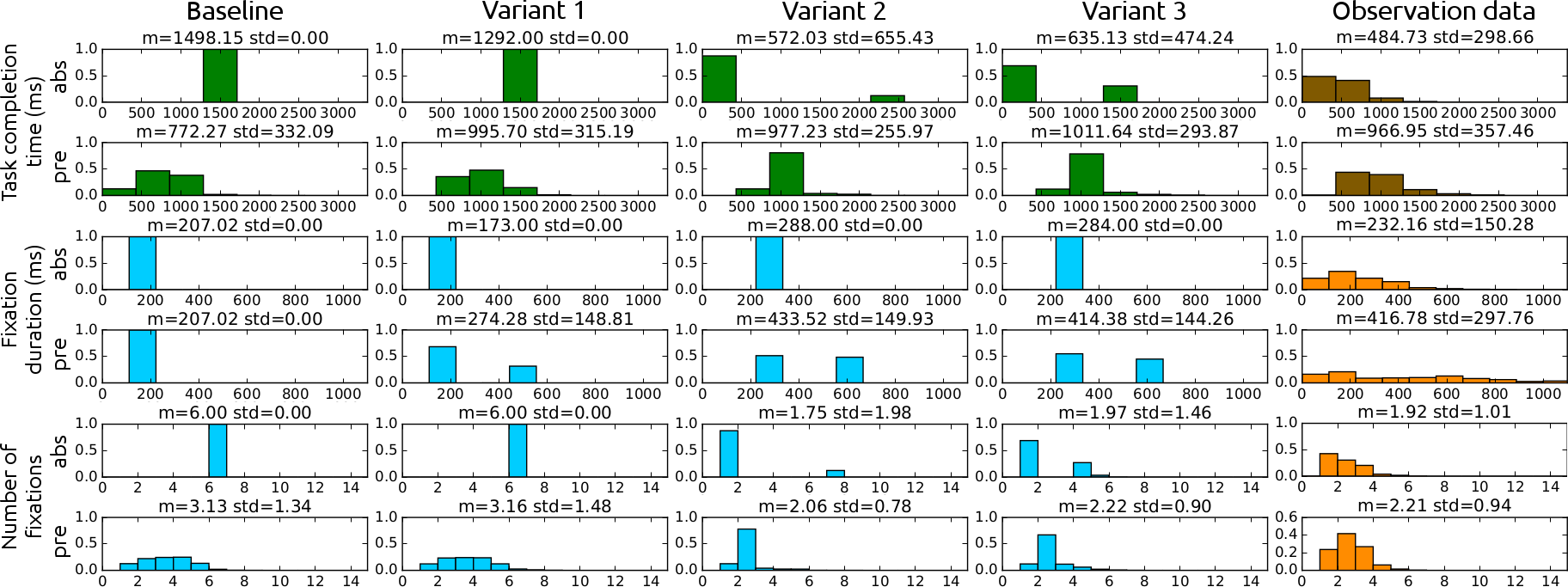}
\caption{
Study 2: ABC exposes how changes to the model (Variants 1-3) result in changes to the predictions when parameter values are conditioned to empirical data.
\emph{Baseline}: Same model as in Study 1, but now conditioned on observed behavior in both target conditions (absent, present) at the same time.
\emph{Variant 1}: Selection delay feature added to baseline (parameter $d_{sel}$).
\emph{Variant 2}: Menu recall feature added to Variant 1 (parameter $p_{rec}$).
\emph{Variant 3}: Peripheral vision feature added to Variant 2 (parameter $p_{sem}$).
\emph{Observation data}: Same as in Study 1.
Reported results are with the MAP parameter values.
\emph{Color coding:} Same as in Figure~\ref{fig:exp1-results-1}.
\emph{Abs}: Data from when target was absent from the menu.
\emph{Pre}: Data from when target was present in the menu.
}
\label{fig:exp2-results-1}
\end{figure*}

The predictions made by the different models, compared to the observation data, are visualized in Figure~\ref{fig:exp2-results-1}.
With increased model complexity, we also see increasing agreement of the predictions with the observation data.
This is partly expected, as more complex models are in general able to fit any dataset better.
However, with the use of priors, we are able to regularize the parameter values to reasonable ranges, and thus avoid over-fitting the models to the data.
 
The baseline model was not able to predict the behavior of the user very well on many of the variables.
The MAP value for $f_{dur}$ (fixation duration) was 210~ms.
The TCTs predicted by the baseline model were [1500~ms (abs), 770~ms (pre)], whereas the ground truth was [490~ms (abs), 970~ms (pre)].
The predicted fixation duration was 210~ms, which is still reasonable, although on the low side, compared to the observed means [230~ms (abs), 420~ms (pre)].
Furthermore, the predicted number of fixations on items on the menu was [6.0 (abs), 3.1 (pre)], whereas the users only performed [1.9 (abs), 2.2 (pre)] fixations.

Variant 1 improved predictions over the baseline.
The MAP value for normal $f_{dur}$ was 170~ms and for $d_{sel}$ (selection delay) 320~ms.
The predicted TCTs were [1300~ms (abs), 1000~ms (pre)], which is already a very reasonable estimate when target is present, although still far from the truth when the target is absent.
The predicted fixation durations (now with the selection delay factored in) were [170~ms (abs), 270~ms (pre)], which is an improvement over the baseline in the present condition, but not on the target absent condition.
The predicted numbers of fixations were nearly identical to baseline.

Variant 2 again improved predictions over both the baseline and Variant 1.
The MAP value for $f_{dur}$ was 290~ms, for $d_{sel}$ was 300~ms, and for $p_{rec}$ (probability of recall) was 87~\%.
The predicted TCTs were [570~ms (abs), 980~ms (pre)], which is the first time we have been able to predict a lower TCT for the target absent case.
However, the variation in TCT when target is absent is quite large;
the predicted standard deviation was 660~ms, whereas the ground truth was 300~ms.
The predicted fixation durations were [290~ms (abs), 430~ms (pre)], which is already close to the ground truth in the target present condition.
The predicted numbers of fixations were [1.8 (abs), 2.1 (pre)], which is a considerable improvement over previous estimates.

Variant 3 provided still slight improvements over previous results.
The MAP value for $f_{dur}$ was 280~ms, for $d_{sel}$ was 290~ms, for $p_{rec}$ was 69~\%, and for $p_{sem}$ (the probability of observing the semantic similarity with peripheral vision) was 93~\%.
The predicted TCTs were [640~ms (abs), 1000~ms (pre)], which is slightly further from the observations than with Variant 2.
However, the variation in the distributions is closer to observed values than with Variant 2 (the discrepancy measure led ABC to minimize both the difference in mean and in standard deviation at the same time, details in Appendix).
The predicted fixation durations were similar as with Variant 2.
The predicted numbers of fixations were [2.0 (abs), 2.2 (pre)], which is slightly better than with Variant 2.

We conclude that we were able to fit multiple model variants to the same observation data, and make meaningful comparisons between the different models because of this.
We observed that the quality of the predictions increased when we added our additional assumptions to the model, which was expected as the models became more flexible, but also provided evidence that these features probably reflect actual user behavior as well.
Furthermore, ABC was found useful in hypothesis comparison, as we avoided manually trying out a large number of different parameter values manually to find values that lead to reasonable predictions.

\subsection{Study 3. ABC and Individual Differences}

Most modeling research in HCI aims at understanding general patterns of user behavior. 
However, understanding how individuals differ is important for both theoretical and practical reasons.
On the one hand, even seemingly simple interfaces like input devices show large variability in user behavior. 
On the other hand, adaptive user interfaces and ability-based design rely on differentiating users based on their knowledge and capabilities.

Our final case looks at the problem of individual differences in inverse modeling.
In Study 3 we select a group of users and fit an individual model for each of these users.
We then compare how good predictions these individual models are able to produce, compared to the same model fit with the data from all of the users in the dataset (population level model).

We selected a representative set of 5 users for Study 3.
We first selected all users from the dataset of whom there were 15 or more observations in each menu condition (target absent, present), leaving 11 users.
We then ordered the users based on their difference in TCT to population mean, summed from both menu conditions.
To get a good distribution of different users, for this experiment we selected the users who were the furthest (S8), third most furthest (S5), and fifth most furthest away (S23) from the population mean -- as well as the users who were the closest (S19) and third most closest (S18) to the population mean.

The model we used in this study, for both individual and population level modeling, corresponded to Variant 3 from the previous section.
To simplify the analysis, here we only infer the values of two of the parameters for each user, keeping the rest fixed.
The inferred parameters were $p_{rec}$ and $p_{sem}$.
Based on the Study 2, it seemed to us that there was less variation in $f_{dur}$ and $d_{sel}$, whereas the use of memory and acuity of peripheral vision could plausibly vary more between individuals.
We fixed the value of $f_{dur}$ to 280~ms and $d_{sel}$ to 290~ms, according to the MAP estimate in Study 2.

For each of the selected users, we collected all of the observations of that user from the dataset, and conditioned the parameter values of the individual model for that user on that small dataset.
The parameter values of the population level model were the same as inferred in Study 2 for Variant 3.
The accuracy of the predictions made for each user by their individual model was compared with the predictions made by the population level model.
In the comparison, we considered the predicted TCTs and numbers of fixations at each condition to the observed values, and report the magnitude (absolute value) of the prediction errors.

\subsubsection{Results}

The predicted MAP parameter values are collected in Table~\ref{tab:exp3-table}.
The individual model parameter values deviate around $\pm$10 percentage points from the population level model parameter values, which is a reasonable magnitude for individual variation.

\begin{figure}[!tb]
\centering
\includegraphics[width=0.47 \textwidth,natwidth=870,natheight=271]{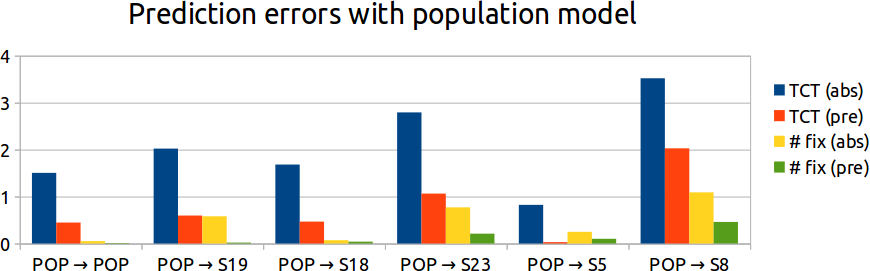}
\caption{
Study 3:
\emph{Leftmost:} Prediction error for population level data with population level model (POP).
\emph{Right:} Prediction errors for individual users (from most to least similar to population mean) with population level model.
Unit of TCT is 100~ms, unit of number of fixations is 1 fixation.
}
\label{fig:exp3-results-1}
\end{figure}

We calculated the magnitude of prediction errors for all of the models by taking the absolute difference in model predicted means and observed data means for each feature.
The prediction errors of the population level model on the population data and on individual user data are shown in Figure~\ref{fig:exp3-results-1}.
Overall, the prediction errors with a population level model tend to be larger for individual users than they are for the whole population.
This shows that population level models that are good for explaining population level dynamics may perform badly when used for explaining subject level dynamics.
Furthermore, as could be expected, prediction errors with a population level model tend to be larger for users who differ more from the population mean.
This presents a clear motivation for developing individual models, as they could help to understand subject level dynamics, especially regarding users who differ from the population mean.

\begin{figure}[!tb]
\centering
\includegraphics[width=0.47 \textwidth,natwidth=870,natheight=270]{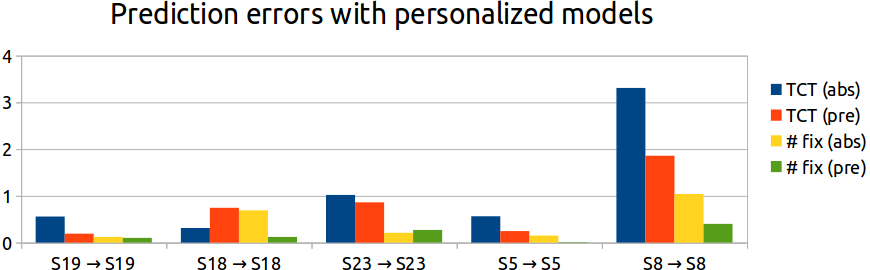}
\caption{
Study 3:
Prediction errors for individual users (from most to least similar to population mean) with models conditioned on observations of the individual user.
Unit of TCT is 100~ms, unit of number of fixations is 1 fixation.
}
\label{fig:exp3-results-2}
\end{figure}

The prediction errors of the individual models on individual user data are shown in Figure~\ref{fig:exp3-results-2}.
Overall we observe a rather consistent quality in the predictions made by the individual user models.
The only exception is user S8, who was the furthest away from the mean.
It is likely that user S8 might have performed the task overall in a very different way from the rest of the users.
For example, the number of fixations taken by this user when target was absent was 3.1, but only 2.7 when the target was present.
This could indicate that the user was unusually careful in examining the menu before declaring that the target was not present.

\begin{figure}[!tb]
\centering
\includegraphics[width=0.47 \textwidth,natwidth=870,natheight=271]{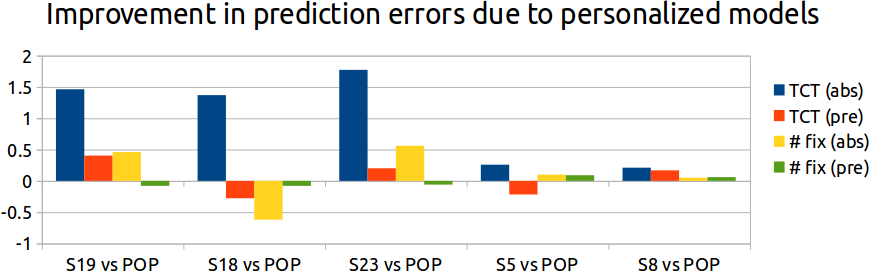}
\caption{
Study 3:
Decrease in prediction error when using individual models (Figure~\ref{fig:exp3-results-2}) instead of the population level model (Figure~\ref{fig:exp3-results-1}) for individual users (from most to least similar to population mean).
Unit of TCT is 100~ms, unit of number of fixations is 1 fixation.
}
\label{fig:exp3-results-3}
\end{figure}

Improvements in prediction error magnitude when changing from population level model to an individual model are shown in Figure~\ref{fig:exp3-results-3}.
The overall trend is that individual user models improve prediction quality, although not always in all parts.
With most users the prediction errors decreased in at least three of the four predicted features.

We conclude that by using ABC we were able to fit CR models to data from individual users, and that the resulting individual models were able to produce better predictions than a population level model fitted to the whole participant pool.
Performing this modeling task would not have been possible with just choosing the values based on literature, as such information tends to only apply for population level models.
On the other hand, choosing the parameter values manually for each user would have required a considerable amount of manual labour, which ABC was able to automate. 
Moreover, inverse modeling helped us expose a behavioral pattern that was not well explained by the model (user S8).

\begin{table}[!t]
\centering
\begin{tabular}{p{1cm}p{1cm}p{1cm}}
\hline
\tabhead{Model} & \tabhead{$p_{rec}$} & \tabhead{$p_{sem}$} \\\hline
S5  & 61~\% & 89~\% \\\hline
S8  & 54~\% & 87~\% \\\hline
S18 & 70~\% & 96~\% \\\hline
S19 & 76~\% & 91~\% \\\hline
S23 & 73~\% & 92~\% \\\hline
POP & 69~\% & 93~\% \\\hline
\end{tabular}
\caption{MAP estimates of parameter values for individual models (S5, S8, S18, S19, S23) and the population level model (POP) in Study 3.}
\label{tab:exp3-table}
\end{table}

\section{Discussion and Conclusion}

We have demonstrated that ABC is applicable for inverse modeling of computationally rational models of complex human behavior based on aggregate behavioural data.

We highlighted advantages ABC has over alternative methods for inferring model parameter values in HCI.
First, the method is applicable for a wide range of models, as it relies on only few assumptions.
Second, the parameter value estimates are conditioned both on the observation data, as well as any prior knowledge the researcher might have of the situation.
This way over-fitting the model to the observation data may be avoided, which could happen if we had only tried to maximize the ability of the model to replicate the data.
Third, the inference process produces a full posterior distribution over the parameter space, instead of only a point estimate, allowing for better analysis of the reliability of the estimates.

In Study 1 we demonstrated that ABC was able to achieve better model fit compared to setting the model parameter value based on literature and manual tuning.
We also identified problems with the existing state-of-the-art model for visual search \cite{chen2015emergence}, related to both the quality of the predictions and convergence issues.

In Study 2 we demonstrated the applicability of ABC in model comparison by fitting four different models to the same dataset and comparing the resulting predictions and inferred model parameter values.
We also proposed improvements to the existing state-of-the-art model, and demonstrated that they resulted in improved quality of predictions.

In Study 3 we demonstrated that with ABC it is possible to fit one of the models from Study 2 to data collected from a single individual, thus creating an individual model.
We further demonstrated that the predictions made by the individual models were better compared to a model fit to a large amount of population-level data.

Together, these contributions help address a substantial problem in understanding interactive behaviour that has been evident in HCI and Human Factors for more than 15 years \cite{kieras2000modern}.
The problem is how to estimate model parameter values given the strategic flexibility of the human cognitive system \cite{Howes2009,kieras2000modern,lewis2014computational}. 
One of the consequences of strategic flexibility has been to make it difficult to test theories of the underlying information processing architecture;
because behaviour that is merely strategic can be mistakenly taken as evidence for one or other architectural theory or set of architectural parameters \cite{Howes2009}.
ABC, and inverse modeling methods in general, addresses this problem by establishing a principled mathematical relationship between the observed behaviour and the model parameter values. 

In the future, inverse modeling might provide a general framework for implementing adaptive interfaces that are able to interpret user behavior so as to determine individual preferences, capabilities, and intentions, rather than merely mapping actions directly to effects.
In summary, we consider ABC to provide ample opportunities for widespread research activity on both HCI applications, and as a core inference methodology for solving the inverse problems arising in research.

\section{Acknowledgments}
This work has been supported by the Academy of Finland (Finnish Centre of Excellence in Computational Inference Research COIN, and grants 294238, 292334) and TEKES (Re:Know).
AH was supported by the EU-funded SPEEDD project (FP7-ICT 619435).
AO was funded by the European Research Council (ERC) under the European Union's Horizon 2020 research and innovation programme (grant agreement 637991).
Computational resources were provided by the Aalto Science IT project.

\section{Appendix: ABC BOLFI Implementation}

We implemented BOLFI in Python with the following details.
We used a Gaussian process (GP) model from the GPy Python library to model the discrepancy.
The kernel was Matern~3/2 with variance 0.01, scale 0.1, and noise variance 0.05.
The first $N_{init}$ sample locations were drawn from the quasi-random Sobol sequence (equal to the number of CPU cores allocated for the job).
The remaining sample locations were decided as follows.
We created a function that computed the lower confidence bound (LCB) for the GP: $LCB(x) = \mu_{GP}(x) - b \sigma_{GP}(x)$.
We used $b =$ 1.0.
For asynchronous parallel sampling, we needed a way to acquire multiple locations that were reasonable, but also sufficiently well apart.
For this purpose we created a function that calculated the sum of radial-basis function kernels that were centered at the locations $P$ currently being sampled: $R(x) = \sum_{p \in P} a \exp((x-p)^2/l)$.
We used $a =$ 0.04, $l =$ 0.04.
The acquisition function for the next sample location was $A(x) = \min_x [LCB(x) + R(x)]$.
Additionally, there was a 10~\% chance of the location being drawn from the prior instead of the acquisition function.

\textbf{Study 1:}
The model was trained with Q-learning for 20 million training episodes, after which we simulated 10,000 episodes for visualizing the behavior predicted by the trained model.
The observation data has the target item absent in 10~\% of the sessions, but in the Chen et al. paper \cite{chen2015emergence} it was assumed that it was absent in 50~\% of the cases.
We tried both splits in training data (10\% and 50\%), but did not find a large overall difference in the results.
In the subsequent experiments, we also used the 10~\% split, as it might remove a possible source of bias.
Our prior for $f_{dur}$ was a truncated Gaussian distribution with mean 300~ms, std 100~ms, min 0~ms, max 600~ms.
The prior was set with the intuition that values between 200~ms and 400~ms should be likely ($\pm$ 1 std), whereas values between 100~ms and 500~ms could still be accepted if the data really supported those values ($\pm$ 2 std).
BOLFI computed discrepancy at 100 locations using 40 CPU cores.
Of the 10,000 simulated episodes, we only used the first 2,500 for calculating the discrepancy.
This was done as it is more sensible to compare datasets of similar size.
Altogether the model fitting took 20~h (in wall-clock time), each individual sample taking 6~h.
The discrepancy was based on the mean and standard deviation of the aggregate task completion time.
It was constructed so that it would fit the mean accurately (L2-penalty) and the standard deviation with lower priority (L1-penalty).
The formula was:

\begin{center}
$d = a \times (mean_{obs} - mean_{sim})^2 + b \times |std_{obs} - std_{sim}|$,
\end{center}

where we used $a = b = 10^{-6}$ for a reasonable scale and the used feature was the aggregate TCT.

\textbf{Study 2:} 
Our prior for $d_{sel}$ was a truncated Gaussian distribution with mean 300~ms, std 300~ms, min 0~ms, max 1000~ms.
300~ms was selected as our initial best guess for the delay, as the second peak in observed fixation duration when target was present (Figure~\ref{fig:exp1-results-1}) was around 600~ms and we thought it likely that the normal fixation duration was around 300~ms.
However, as we had relatively high uncertainty about this, we chose a quite flat prior.
Our prior for $p_{rec}$ and $p_{sem}$ were uniform distributions with [min 0, max 1].
Uninformative priors were used as we were uncertain about the possible true values of these parameters.
The discrepancy was the average of $d(TCT_{pre})$ and $d(TCT_{abs})$.
As the parameter space sizes varied, we chose the number of samples and CPUs for each case separately.
Baseline: 100 samples, 40 CPUs (20 h).
Variant 1: 200 samples, 80 CPUs (16 h).
Variant 2: 400 samples, 80 CPUs (30 h).
Variant 3: 600 samples, 100 CPUs (37 h).

\textbf{Study 3:}
The prior for $p_{rec}$ was a truncated Gaussian distribution with [mean 69~\%, std 20~\%, min 0~\%, max 100~\%].
The prior for $p_{sem}$ was similar but with mean 93~\%.
The priors were based on the knowledge gained from Study 2, and thus centered on the MAP estimate of Variant 3, but were reasonably flat to allow for individual variation.
The discrepancy was the same as in Study 2.
Out of the total 10,000 simulated sessions, we used the first 200 for calculating the discrepancy to match the individual dataset sizes.
For each of the users, we computed 200 samples using 60 CPUs (22 h each).

\bibliographystyle{SIGCHI-new}
\bibliography{references}
\balance

\end{document}